\documentclass{article}
\usepackage{graphicx}
\usepackage{amsmath}

\newcommand{\hs}{\hspace*{5ex}}

\begin{document}
 
\baselineskip 18pt

\begin{center}
{\Large {\bf  Status of Radio and Acoustic Detection of Ultra-High Energy
Cosmic Neutrinos\\ and\\ a Proposal on Reporting Results}}\footnote{
Presented at Nobel Symposium \#129 (Neutrino Physics), 
Enk\"{o}ping, Sweden}

 \vskip5mm David Saltzberg\vskip3mm \mbox{}%
Dept. of Physics and Astronomy, University of California,
Los Angeles (UCLA), 475 Portola Plaza, Los~Angeles, California, 
90095-1547, USA

\bigskip

\begin{abstract}

Neutrino astronomy offers the possibility to perform
extra-galactic observations well beyond the photon absorption
cutoff above $5\times10^{13}$~eV.   
Based on observations of cosmic rays, we already know 
that astrophysical
sources produce particles with at least a million times
more energy than this photon cutoff.  
Once
discovered, either the nature of the sources
themselves or the cross sections of ultra-high
energy neutrinos with terrestrial matter may reveal exotic 
physical processes that are inaccessible to modern accelerators.
Some of these
processes may be due to as-yet unknown physics at the grand
unification scale or beyond.    Neutrino telescopes based on
optical techniques currently operating and
under construction have apertures measured in several
km$^3$-sr.   Radio and acoustic detection techniques have
been demonstrated in laboratory experiments and are currently
used for instrumentation of apertures 10~to 10,000~times
larger than optical techniques 
for neutrinos above $10^{16}$~eV.   I discuss the
status of current and proposed neutrino telescope projects
based on these techniques.    These telescopes have
already ruled out some of the more exotic predictions
for neutrino intensity.   The upcoming generation of 
radio-based and acoustic-based detectors
should be sensitive to cosmic neutrinos above $10^{18}$~eV 
originating through the so-called GZK process.   A comparison
of different neutrino telescopes using a common aperture variable
shows how they are complementary in the trade-off of volume versus
threshold.  I include a proposal for how neutrino telescopes
should report their sensitivities to facilitate direct comparisons
among them and to allow testing of neutrino brightness 
models that appear even  after publication of the experimental results.

\end{abstract}
\end{center}
\newpage

\section*{1. Cosmic Neutrinos for Astrophysicists and Particle Physicists}
 
\hs Both astrophysical and particle-physics questions
drive the quest to detect high-energy ($>10^{12}$~eV) cosmic neutrinos.
To date, the only two detected neutrino sources,
albeit at lower energies ($5-50$~MeV), have illustrated how such observations 
can significantly impact
both fields.   First, the observation of the
expected flux of neutrinos from the Sun provided
astrophysicists with the first 
direct view of the central engine, not just surface, 
of the Sun and confirmed the predictions of the standard solar model.
For the particle physicists, measurements of
the flavor content of this neutrino flux yielded two compelling
facts about neutrinos:  neutrinos have mass and
the weak neutrino eigenstates are not their
mass eigenstates.   Second, observations of neutrinos
emitted from SN1987a in the Large Magellanic Cloud found the
predicted prompt neutrino production from the cataclysmic
astrophysical event itself, estimated to be of order 99\%
of the energy release.  For particle
physicists, the lack of dispersion of the arrival
times of these neutrinos over such a long baseline yielded
direct limits on the neutrino mass complementary to
the contemporary laboratory measurements, and subject to 
entirely different systematic uncertainties.   
To date, the observation of
every cosmic neutrino source has resulted in a Nobel Prize.
   
   Since neutrino cross sections increase with energy, while
backgrounds due to atmospheric neutrino production decrease,
most searches for other sources of cosmic neutrinos operate at
high energies.  Potential sources
for these neutrinos include black holes accreting matter
in active galactic nuclei (AGN)~\cite{agn} 
and in fireballs associated with
gamma-ray bursts (GRB).  Other considered sources include
so-called topological defects (TD)~\cite{td}, remnants of the early
universe which might produce particles with
masses comparable to various unification scales such as
$10^{25-28}$~eV, that in turn decay to particles producing 
neutrinos.  Observations of AGN and GRB neutrino sources
would likely yield insight into astrophysical processes 
responsible for accelerating the highest energy cosmic
rays.   Alternatively, observation of TD neutrinos
could provide insight into elementary particle physics 
unobtainable through direct production at 
terrestrial accelerators.

   An additional, well constrained theoretically, source of cosmic neutrinos
is produced by the GZK process,
the inelastic collision of cosmic ray protons 
with energies above $10^{19.5}$~eV 
with the 2.7~K microwave
background photons.  The collisions occur with
a mean free path 
short compared to 
cosmological distance scales~\cite{gzkcr}
(of order $4$~Mpc for protons with energy
$6\times 10^{20}$~eV).
The energy threshold for protons to interact this way, $10^{19.5}$~eV,
corresponds to the center-of-mass threshold energy for pion production,
about 1200~MeV.
The pions' decay chain produces the so-called 
GZK neutrinos with energies above $10^{17}$~eV~\cite{gzknus1},
sometimes called the ``guaranteed'' neutrinos.
Assuming our local intensity of such protons is no different
than the rest of the universe, the intensity of neutrinos is predicted
with only an order-of-magnitude uncertainty~\cite{gzknus2}.
The astrophysical interest in these neutrinos
is not that they allow astronomers to look deep into the
core of an astrophysical object, but rather that significant
deviation from the predicted level would point to a lack
of understanding of the relevant energy flows.
Note that these predictions are based on proton intensities
at $10^{18}$~and $10^{19}$~eV, and do not use measurements
at or above the GZK cutoff at $10^{19.5}$~eV.
The only significant loophole is that
the sources themselves could have an independent internal cutoff 
at $10^{19.5}$~eV, which would  be a cruel coincidence.

   The observation of any source of cosmic neutrinos would also provide
a unique laboratory for particle physicists.    Neutrino
interactions above $10^{18}$~eV would probe the electroweak
and other interactions at center-of-mass energies beyond those
obtainable with modern particle accelerators.   Because
the standard-model neutrino cross sections are so small,
deviations from the standard model due to
micro-black hole production or other
non-perturbative effects~\cite{microbh} could
make them dramatically larger, even by several orders of magnitude.
The value of the cross section could
be studied by the absorption of neutrinos from
a cosmic source through
the Earth's crust and perhaps atmosphere~\cite{diffabs}.
By contrast, even though cosmic-ray proton or ion interactions
probe similarly large center-of-mass energies, their interactions
are hadronic, so these additional channels
would produce only a small change to their already large
cross sections.

\section*{2. Quantifying Cosmic Neutrino Detection}

\hs In neutrino astronomy, as in other types of astronomy, it
is important to distinguish the concepts of {\it flux,
intensity, and brightness}.   I will
use these terms as they appear in radiative transfer
theory~\cite{rybicki}, with the modification that I will
consider the transfer of number of neutrinos rather than heat.
{\it Energy} here will refer to the individual energy of
each neutrino, $E$, in analogy to the photon frequency,
$\nu$, in radiative transfer theory.    This convention
is consistent with the comprehensive review by
Learned and Mannheim~\cite{learnedmannheim}.

In Table~\ref{defns}, I give the names and units used
for the relevant variables.   Detectors measure
a number of neutrinos, $N_\nu$, through an area, $A$, 
for a given solid angle, $\Delta\Omega$, during
a livetime, $\Delta t$, perhaps with some event-by-event
energy information.  For these proceedings
I assume that sources produce a neutrino output that
is constant in time; however, based on high energy 
gamma-ray observations of AGN, observers should
be prepared to see flaring objects.   For diffuse
sources, the relevant variables are intensity, $I$,
and brightness, $I_E$.   Often these are
denoted as $d^3N/dAd\Omega dt$ and
$d^4N/dAd\Omega dtdE$, respectively.
For point sources,
the relevant variables are the flux variables given
in the table, which
I will not consider further here.  Sometimes ``intensity''
is referred to as ``diffuse flux''.  Brightness from
isotropic models is commonly abbreviated as $dN/dE$, leaving
the differentials with respect to area and solid angle as
implied, but typically indicated in the units.

The reader should be careful to note the distinction
between intensity and brightness.   Neutrino intensity
is calculated or measured as an integral over energy, so it
is not itself a function of energy.  The reader will
be reminded of the extent of the integral by denoting the
intensity as ``$I|_{E_1}^{E_2}$''.
In practice, theorists typically provide
predictions on neutrino brightness, $I_E$,
while experimentalists
provide sensitivity to neutrino intensities, $I|_{E_1}^{E_2}$.

   For quantifying detection and sensitivities I consider that
there exists an unknown brightness of ultra-high energy
neutrinos which might depend on $E$, $\theta$, and~$\phi$.
Assuming the brightness is isotropic, it depends only
on $E$.   
The brightness $I_E$ corresponds to no physical
observable and for the experimentalist
exists only under an integral sign:
\begin{equation}
I|_{E_1}^{E_2} = \int_{E_1}^{E_2} I_E(E) dE,
\end{equation}
where $I|_{E_1}^{E_2}$ is the number of neutrinos per area per steradian
per time with energy between $E_1$ and $E_2$.  (For this
reason, brightness may be thought of as an ``intensity density'',
in analogy with the use of
``flux density'' for $F_E$, but this is not the common term.)

   The properties of a neutrino telescope determine the
number of events it will detect.   First, note that a simple
flat particle counter with 100\% efficiency is typically
described by 
the area, $A$, and solid angle coverage, $\Delta \Omega$.
In general, one only considers the product $\alpha\equiv [A\Delta\Omega]$ vs.
energy which usually cannot be uniquely factored
(For example, even for a simple flat paddle that is sensitive
to particles crossing in either direction, $\alpha=[A\Delta\Omega]$
does not unambiguously factor.  It has $\alpha=A\times 2\pi$
not $A \times 4\pi$ due to projection effects.  This 
relative factor
of one-half can be absorbed either into an effective area or
effective solid angle.)   A neutrino telescope typically
has a very low efficiency for neutrinos but still can
be thought of as having an equivalent area with 100\%
efficiency,
which is typically a strong function of neutrino energy.
Thus during a time interval $\Delta t$, a telescope
observe $N_\nu$ events in the interval $[E_1, E_2]$:
\begin{equation}
N_\nu = [\sum_{i=1}^{3} \int_{E_1}^{E_2} \alpha^i(E) \ 
I_E^i(E) \ dE] \ \Delta t,
\label{nevents}
\end{equation}
where $i$ is a sum over the three neutrino species.   This
may be considered the defining equation of $\alpha(E)$ for
each neutrino species.

If the number of expected background events is much less
than one, one may compare the sensitivities 
of neutrino telescopes directly using
$[A \Delta \Omega](E) \Delta t$.  Note that the comparison
must be a function of neutrino energy.   However,
in comparing experiments, one sometimes needs to take into
account the expected backgrounds to properly compare
the sensitivities.  Here I define a quantity~\cite{greens}
\begin{equation}
\mbox{discovery aperture}=[A \Delta \Omega](E) \Delta t/N_{3\sigma}, 
\end{equation}
where $N_{3\sigma}$
is the number of events the experiment would need to
see a $3\sigma$ excess over background.   (Note that although high-energy gamma
ray detectors often use $5\sigma$ as their standard, they are
usually looking for point sources in many places on the sky.
Here one is concerned with detection of an isotropic intensity
for which there are many times fewer trials.)   For a
background-free experiment, take $N_{3\sigma}=1$.
  
  Experiments commonly report an effective water-equivalent
volume times steradians, $[V^{w.e.}\Delta\Omega]$ rather
than $[A\Delta\Omega]$.  The variable $[V^{w.e.}\Delta \Omega]$
is defined so that the two descriptions become equivalent
in the thin-target approximation via
\begin{equation}
[V^{w.e.}\Delta \Omega](E)\equiv [A\Delta\Omega](E)\times \ell_{int}^{H_2O},
\label{volsr}
\end{equation}
where $\ell_{int}^{H_2O}$ is the mean interaction length for
the neutrino in liquid water at that energy.  Specifically,
$\ell_{int}^{H_2O}=m_{amu}/(\sigma \rho^{H_2O})$, where $\rho^{H_2O}$ is
the density of water, $m_{amu}$ is the atomic mass unit
and $\sigma$ is the neutrino cross section per nucleon~\cite{gqrs}.
In the rare case, typically at very high energies, that 
the active detector
medium itself is large enough to produce 
some neutrino shielding, it would be less 
ambiguous to report just $[A\Delta\Omega](E)$.

There are
two reasons why the variable $[V^{w.e.}\Delta\Omega]$ is sometimes used.
First, the detectors physically occupy volumes; so one can
attempt to more sensibly factor this into effective
$V^{w.e.}_{eff}$ and $\Delta\Omega_{eff}$.
Second, this value does not appear to depend on the
neutrino cross section.     However, for some high energy
telescopes a large portion of their aperture comes from
events outside the nominally instrumented volume.
Also since neutrino
telescopes above $10^{14}$~eV need to include upstream
shielding of neutrinos by the Earth, the independence
is far from complete.  In practice, each telescope
may have a very different dependence on the value
of neutrino cross section.   Since one usually needs
the experiments' own Monte Carlo simulations to know
the variance of the sensitivity with cross section,
neutrino telescope collaborations would do better to
publish either $[A\Delta\Omega](E)$ or $[V^{w.e}\Delta\Omega](E)$
for values of 75\%, 100\% and 125\% of the nominal
cross sections used.   

   Looking at the typical neutrino
brightness predictions~\cite{agn},~\cite{td},~\cite{gzknus2}
in the context of equations~\ref{nevents} and~\ref{volsr},
it becomes apparent that for neutrino energies $10^{13}$~eV to 
$10^{15}$~eV, one needs detectors with $[V^{w.e.}\Delta\Omega]$ 
of order
$1-10$~km$^3$-sr.    Already these sizes demand instrumenting
large natural volumes since they are larger than anything
that can be reasonably manufactured.  
Several telescopes are currently running or
under construction that employ optical techniques in the
clearest media known:  Antarctic ice (Amanda, IceCube)
and deep water (Baikal, Antares, Nestor, and Nemo).   The limiting size of
such detectors is determined by the effective attenuation length of light
(including scattering), 50-100~m in these media.  This length 
determines spacing and the practicality
of deployment, yielding detectors that may 
achieve apertures up to $10$~km$^{3}$-sr.

  For energies above $10^{18}$~eV, 
one can attempt to detect the GZK neutrinos whose
brightness is fairly well understood.
The desired size increases to at least 1,000~km$^{3}$-sr.
So while the standard optical detection techniques
are well suited to energies up to $10^{15}$~eV,
larger volumes are needed at higher energies.   As will be described
in Sections~3 and~4, neutrino-induced showers also produce radio emission up
to tens of~GHz and acoustic emission to hundreds of~kHz.
Since in some materials
the attenuation length for these emissions may exceed a kilometer,
several groups are exploiting these techniques to achieve the extremely
large apertures.  
(Additional information 
may also be found in another recent review~\cite{nahnhauer}.)
Two other techniques also offer the
possibility of such larger apertures:  atmospheric observations from space
and exploiting mountain ranges and the finite lifetime of the
$\tau$ lepton; these will be discussed in Section~5.  Based on
some lessons learned from this exercise, I present
a proposal on reporting results in Section~6.  

\section*{3. Radio Detection}

\hs When a neutrino interacts in material, it produces a shower
of relativistic particles.  For energies above $10^{15}$~eV, about
20\% of the neutrino energy appears as a relativistic hadronic shower
through the ``inelasticity'' of the neutrino interaction with matter.  The
hadronic shower eventually becomes electromagnetic (electrons, positrons,
and photons) through production of 
$\pi^0$ mesons which quickly decay to photons. 
For charged-current interactions, which are 70\%
of the interactions at these energies, a charged lepton, $e$, $\mu$,
or $\tau$ is also produced with the remaining energy, each 
of which eventually deposits some 
electromagnetic energy as well.   These particles are typically
moving faster than the speed of light in the material,
$c/n_{refr}$, and produce the optical Cherenkov radiation that is the
basis for many neutrino telescopes.

  In the early 1960's, Gurgen Askaryan realized~\cite{askradio}
that a strong radio component
(up to $\sim$ 10~GHz, lasting of order 1~ns) of Cherenkov
emission would be produced as well.
For interactions in matter, a $(10-30)$\% charge excess of electrons
over positrons will develop since the target material contains
electrons at rest and no positrons.   For wavelengths longer than
the lateral size of the shower (a few centimeters in radius)
the emission becomes coherent.  Unlike the optical radiation,
which is incoherent, the power of radio emission grows as
the square of the energy of the shower.  At $10^{18}$~eV, about $10^8$ 
electrons radiate
in phase, producing a large electromagnetic pulse 
(of order $500$~V/m at 1~m.)
Askaryan's calculations were subsequently
confirmed by independent groups using
modern shower simulations based on EGS~\cite{zhscalcs} and 
with  GEANT~\cite{geantcalcs}, which are now in basic agreement.
The predicted characteristics and intensity of
emission were further demonstrated in a series of experiments
at SLAC~\cite{slacexps}.    The square-law dependence of
the power in incident neutrino energy, which helps for
high neutrino energies, unfortunately also implies
a relatively high threshold due to the presence of 
irreducible thermal noise in any detection medium.  Thresholds
as low as $10^{17}$~eV are achieved with this technique, perhaps
extendible as low as $10^{16}$~eV.
In his original work, 
Askaryan presciently pointed out that detectors could
be based on large ice sheets, natural salt formations and
the lunar surface material which to date still form the basis
of this technique.

\begin{figure}[!hbp]
\begin{center}
\includegraphics[clip=true,scale=0.9]{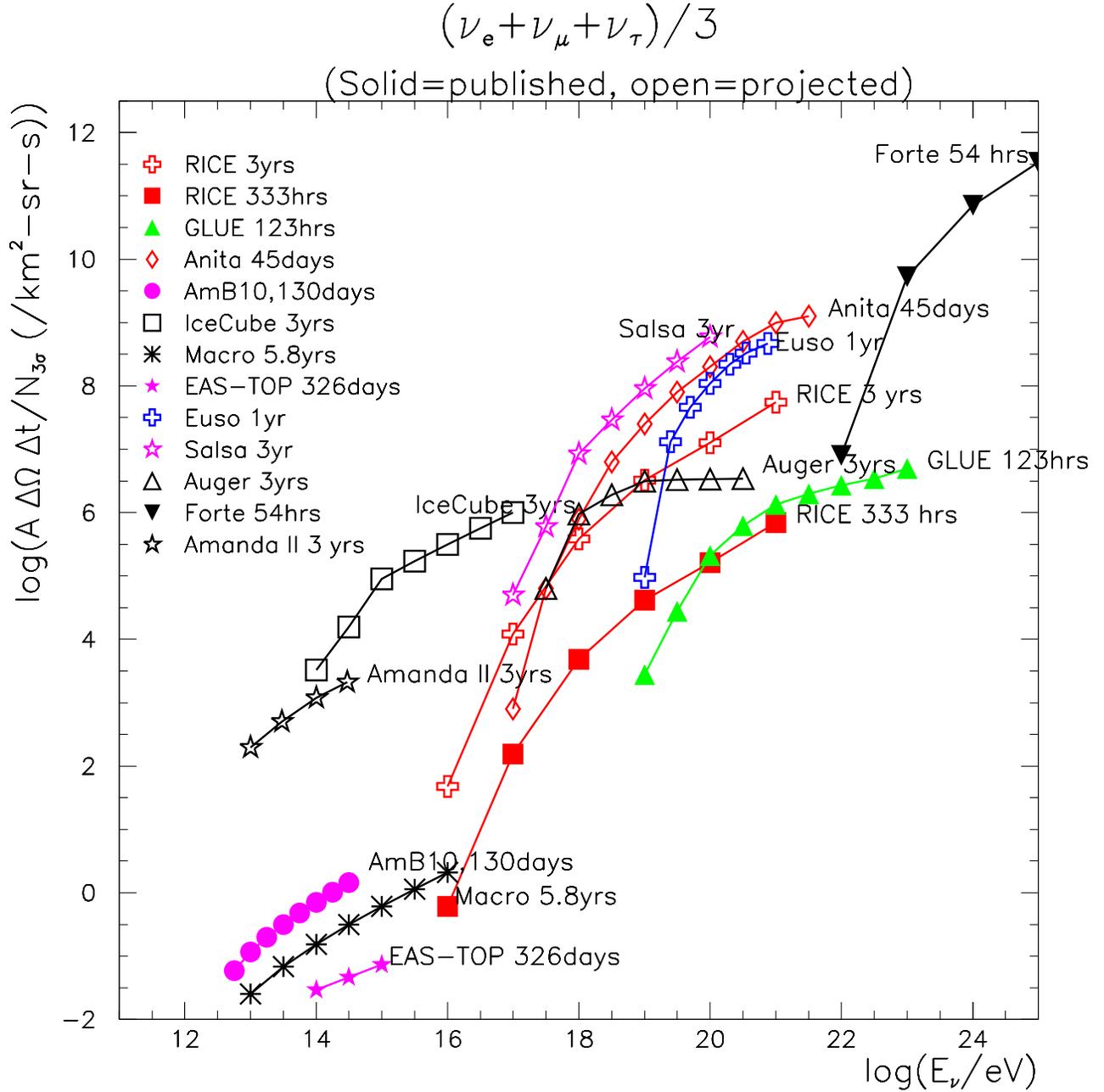}
\bigskip
\caption{
My attempt to convert several telescopes' sensitivity to a common
discovery aperture,
$[A \Delta \Omega]\Delta t/N_{3\sigma}$, as defined
in the text.   The quantity presented here
is the average over all three neutrinos species, for illustrative
purposes but an
analysis breaking it down by flavor, such as in Ref.~\cite{conversion}, 
may also be done.
The values presented
here should be taken as representative, awaiting official numbers
from the collaborations themselves, so comparisons should be
taken  to within a $\log_{10}(2)=0.3$ ``grain-of-salt''
on the vertical and horizontal scales.
Current results are presented with solid symbols, projected
results are with open symbols.  Note that some projected telescopes
are funded and under construction, while others  are in the conceptual
stage.  Times are live time.  CAVEAT:  Because this plot presents
the average over three neutrino flavors, telescopes that obtain
most of their sensitivity from one flavor (Amanda, Auger) would
do better relative to other experiments in pure-flavor plots.
When space allows, three separate plots for the individual
flavors would be better.
\label{apertures}}
\end{center}
\end{figure}

The RICE~\cite{rice} telescope consists of 20 dipole
antennas sensitive after filtering from 200~to 500~MHz placed on the same
strings as the Amanda neutrino telescope at the South Pole.
The array comprises a $200\times 200\times 200$~m array.  The
ice on the Antarctic plateau has
exceptionally long attenuation lengths, in excess of 1000~m,
as extracted from airborne ground-penetrating
radar and surface measurements in Greenland, Iceland,
laboratory measurements and a new South Pole 
measurement~\cite{attens}.  Hence,
the sensitive volume
is much larger than the array itself, achieving above
10~km$^3$-sr above $10^{18}$~eV with nearly full-time duty
cycle.   RICE has 
placed limits
on cosmic neutrinos with energies above $10^{16}$~eV based
on a 3-year dataset and data taking is ongoing.
The RICE aperture
is compared to current and expected techniques in 
Figure~\ref{apertures} and
Table~\ref{predictions}.  Studies of an extension
of this technique, X-RICE~\cite{xrice},
to holes spread over an area up to 10$^{4}$~km$^2$ in Antarctica
could be sensitive to tens to hundreds of GZK neutrinos per year.

Based on an idea of Zheleznyk and Dagkesamanskii~\cite{zheldag},
the emission from neutrino interactions in the outer $10-20$~m
of the Moon's surface is detectable above its thermal
black-body emission for energies above $10^{20}$~eV by
terrestrial radio telescopes.   A 12-hour search
using the Parkes 64~m dish
in Australia~\cite{hankins}
did not see any events but had significant
radio-frequency interference.  A search using
radio telescopes at the Goldstone tracking station 
(GLUE)~\cite{glue} used two antennas in coincidence to
eliminate terrestrial background.  GLUE did not see any
events after 123~hours of observations with
a threshold of $10^{20}$~eV and achieving
$[V^{w.e.}\Delta \Omega]$ of order 500~km$^{3}$-sr, 
although unlike RICE, the duty cycle 
due to available telescope time corresponded to 
only $5\times10^{-3}$.
The GLUE aperture
is compared to current and expected techniques in 
Figure~\ref{apertures} and
Table~\ref{predictions}.  Future attempts using this
technique are underway at the Kalyazin radiotelescope~\cite{kalyazin}.

The FORTE experiment~\cite{forte} used a space-based platform
with a log-periodic antenna array 
designed for VHF lightning observations to achieve a total
of 2.3~days integrated observation of Greenland ice.  (Its
orbit did not provide a view of Antarctic ice.)   Since the radio
emission travels through the ionosphere,  there is a characteristic
frequency-dependent delay which allowed suppression of backgrounds.
Because
FORTE operated at a lower frequency and bandwidth than other experiments
(22~MHz BW from 30~to 300~MHz) 
it suffered from a higher threshold, $10^{22}$~eV.  However 
at these frequencies
the Cherenkov radio emission diffracts due to the finite
length of the shower and thereby fills a large solid angle.
FORTE thereby achieves an impressive 
$[V^{w.e.}\Delta \Omega]\sim 100,000$~km$^3$-sr.
Its duty cycle was $\sim 3\times 10^{-3}$.
The FORTE aperture
is compared to current and expected techniques in 
Figure~\ref{apertures}.

The ANITA~\cite{anita}
detector combines many of the most attractive features of the above
experiments into one project designed to be sensitive to
the GZK neutrinos.  ANITA will be a balloon-borne payload of
dual-ridged gain horn antennas (200-1200~MHz after filtering) flying
for a total of 60 days beginning
in the 2006-07 season.   At an altitude of 37~km above the
Antarctic ice, ANITA will have a threshold of approximately
$10^{17}$~eV and with an instantaneous view of 1.5 million
km$^{3}$ of ice, yielding an
aperture of $\sim$20,000 km$^{3}$-sr.   Assuming an
18-day flight per year, the duty cycle is an annualized 0.05.   
A test  flight
in the 2003-04 season, ANITA-lite~\cite{anita}, 
indicated a sufficiently radio-quiet
environment and successfully demonstrated
the essential in-flight systems.   This engineering flight worked
well enough that a new limit may be extracted from these data.
The ANITA aperture
is compared to other expected results in 
Figure~\ref{apertures} and
Table~\ref{predictions}.

A complementary approach that is still in
the conceptual stage is to instrument large natural
salt formations with antennas much in the style of
the RICE detector.       Salt domes which are large and dry offer 
attenuation lengths in excess of a few hundred meters
which, accounting for the increased
density, correspond to nearly a kilometer water-equivalent
attenuation length.   Example detectors include the
SND~\cite{snd}, SALSA~\cite{salsa} and~ZESANA~\cite{zesana}
concepts which could
instrument many tens of cubic kilometers of salt.
Because the antennas are closer to the events than ANITA, the threshold 
could be an order-of-magnitude lower.  Even though
$[V^{w.e.}\Delta\Omega]$ is comparable or even lower than
ANITA,  because the duty cycle
is much larger, the ultimate sensitivity of salt-based detectors
is the largest of any considered and has
the lowest threshold.  In addition, there would be
closer to full-sky coverage and a larger duty cycle for 
flaring sources.   Because salt-domes are generally covered
by an overburden of rock, they are naturally shielded from
man-made electromagnetic sources on or above the 
Earth's surface.   The SALSA aperture
is compared to other expected results in 
Figure~\ref{apertures} and
Table~\ref{predictions}.

\section*{4. Acoustic Detection}

\hs In addition to the radio detection idea, Askaryan described
how ultra-high energy neutrino interactions could be detected
by underwater acoustic techniques~\cite{askacoust}.   A 
$10^{21}$~eV neutrino shower under water deposits 150~J of
heat through ionization in a highly localized region.  As a result, a 
hydrothermal pressure impulse
lasting of order 10~$\mu$s propagates outward as a thin 
pancake. 
Verification of the production
of this pulse was performed at accelerators at Brookhaven,
Uppsala, and ITEP~\cite{accelacoust}.

Noise sources in the ocean
include turbulence, surface waves, 
wind, oceanic traffic, precipitation, biological systems 
and thermal noise.  The noise spectral density goes through a minimum
in the 10~kHz range which is also where the power spectral density
of the neutrino-induced pressure pulse would peak.
Work is underway to characterize the noise
environment in solids such as Antarctic ice and salt domes.

Over the years, several detectors have been proposed, generally
using hydrophone arrays in the deep ocean.
Considerable interest has been renewed lately. R\&D for
acoustic sensors and site selection was discussed recently at
a workshop dedicated to the acoustic technique~\cite{acousticworkshop}
and can be found among the slides.  
Hydrophone arrays near
Kamchatka (SADCO)~\cite{kamchatka},
Rona (U.K.)~\cite{rona}, and TREMAIL near the Antares
site have been used for preliminary environmental measurements
and testing of reconstruction techniques.
At the acoustic workshop, a large number of groups showed
significant and broad progress in development of hydrophones and
pre-amplifiers for the Nemo, Antares, and Lake Baikal 
sites.   
The existing Lake Baikal sensors may even show a hint of excess consistent
with an acoustical neutrino
signal~\cite{budnev} and their data-taking continues.
Detector development for solid media such as Antarctic ice
and salt domes is also underway~\cite{solidacoust}.
Well-understood calibrators for
solid and liquid acoustic detectors are critical; methods under
development include submerged implosions and noise from airplanes.
Until recently, sensitivity
estimates have been largely analytic, but 
Monte Carlo simulation
tools for acoustic detection are being developed~\cite{watersniess} 
that will aid future design and sensitivity estimates.

Here I discuss one acoustic-based telescope
in detail, the SAUND detector~\cite{saund},
since it has been carried through to completion, including a
publication.
Located in the deep sea off of the Bahamas, the SAUND collaboration
instrumented 7 hydrophones arranged 
in a star pattern using a subset
of the U.S. Navy's AUTEC array in the Bahamas to detect
the impulse in frequency bands $7-50$~kHz.   The hydrophones
are deep (1.6~km) 
and sheltered by several islands and shoals with
little shipping traffic.
The observed
noise floor, consistent with $\nu^{-1.7}$ Knudsen noise set
the threshold at $\sim 10^{21}$~eV but could be reduced by finding
or building a hydrophone array with closer spacing.
Imploding light bulbs at various distances and depths
tested the
event reconstruction
and were consistent with attenuation lengths of 500 to 1000~m.   Refraction
effects become significant at 1000~m and beyond.
The collaboration ran a physics run for 195 days.  Although the 
$[V^{w.e.}\Delta \Omega]$ of about  100~km$^{3}$-sr is 
still too small for these energies, it should be noted
that SAUND's geometry was not optimal for the pancake-like shape
of the acoustic pulse.  There
is hope that arrays with optimized geometry
and also solids such as large salt domes could provide 
one or two 
orders of magnitude more aperture through larger
signal and lower backgrounds.     
The SAUND group is preparing
to install SAUND-2 to instrument 1500~km$^{3}$ of sea water~\cite{saund2}.

A hybrid detector combining both radio and acoustic techniques
holds some promise.  Due to the extreme difference in the velocities
of propagation of radio versus sound waves, and their different
polarization properties, a simultaneous detection of the
same event using the two techniques might yield interesting
information.  However, if the thresholds and sizes of the two arrays
are significantly different, they would be unlikely to make
any simultaneous observations.

Considerable new activity has begun in the last couple of years
in the field of acoustic detection techniques for neutrino astronomy
after a relatively quiet two decades.   The 
radio and acoustic  fields are now so rich that future summaries
ought to be divided among two speakers to do the fields justice.

\section*{5. Other neutrino telescope techniques}
\label{other}

\hs We wish to compare the radio and acoustic techniques to other neutrino
telescopes that have run or will turn on that are sensitive to neutrinos
above $10^{12}$~eV.   
Often it is difficult based on the published materials
to convert the reported limits into the sensitivity curve 
(discovery aperture) $[A\Delta\Omega]\Delta t/N_{3\sigma}$ versus energy. 
I have converted their limits
in a hopefully reasonable way~\cite{conversion} and
show the comparisons in Figure~\ref{apertures}.  Numbers of
expected events based on these estimates and equation~\ref{nevents} are
shown in Table~\ref{predictions}.

In the $10^{12}$ to $10^{15}$~eV regime, several detectors have looked
for muons produced by $\nu_\mu$ charged-current neutrino interactions
or pion decay  in the material below them. 
MACRO was a small detector but ran for nearly 6 
live years and set limits
on the neutrino intensity between $10^{13}$~and $10^{16}$~eV.
Recently Amanda-B10, which is a much larger detector,
set a stronger limit  on the $\nu_\mu$ intensity between 
$5\times 10^{13}$~and $5\times 10^{14}$~eV
in less time.   The Lake Baikal
neutrino telescope~\cite{baikal}  reports limits with a threshold around
$10^{13}$~eV for $\nu_\mu$ interactions.
Using a different technique EAS-TOP~\cite{eastop}
ran for 326 days and reported limits for $E$ between $10^{14}$~and 
$10^{15}$~eV based on the rate of horizontal
extensive air showers.
The Amanda-II detector~\cite{am2}, consisting of 19~strings will set even
strong bounds.
As shown in the figure, 
from $10^{14}$~to $10^{17}$~eV the IceCube~\cite{icecube} detector will
greatly increase the current sensitivity.   
Telescopes based in the Mediterranean
(Antares, Nemo and Nestor) ~\cite{medit}, using sea water rather
than Antarctic ice 
will both overlap and extend the sky coverage of IceCube.   
The sensitivities of a few representative optical-based
telescopes are compared with the radio and acoustic techniques in
Figure~\ref{apertures}.

Above about $10^{16}$~eV, a few other techniques besides radio and
acoustic offer the possibility of extending beyond 10~km$^{3}$-sr.
The Auger collaboration would be sensitive to $\nu_\tau$ interactions
in nearby mountains; because due to the finite lifetime of the $\tau$
lepton, they would be able to detect its decay with 
$[V^{w.e.}\Delta\Omega]$ as large as 10~km$^{3}$-sr above
$10^{18}$~eV.    This has the advantage that it can be done
parasitically with an existing experiment.
The EUSO collaboration~\cite{euso} showed how observing large
volumes of the atmosphere from an orbiting platform could 
be sensitive to
$[V^{w.e.}\Delta \Omega]$ of order 100~km$^{3}$-sr with a high
duty cycle, subject to the detector being launched.   
Still, as shown in the figure, the apertures
achievable with radio detection are larger.   Radio-based
detectors in embedded in salt or ice, with their
relatively low threshold, large duty cycle and large sky coverage
may eventually be the most versatile ultra-high energy
radio-based neutrino detector.

\section*{6. Proposal for how neutrino telescopes should
report results}

\hs Because I do not have access to each collaboration's internal
Monte Carlo simulations and/or calculations I took some liberties
of interpretation in converting the reported numbers into a common
framework.  The most common problem was that 
some limits already included
the neutrino cross section and 
even particular models for neutrino brightness
that were hard to untangle.   Sometimes ``effective area'' or
``effective volume'' was quoted without stating  for which solid
angle it was defined.   Sometimes the ``effective volume'' did not
state if it was for water-equivalent or for the actual medium's density.
Sometimes it was not even clear if deadtimes
were already included in the reported livetime or absorbed into the limit
as an efficiency.  As a result of this effort,
I propose here how neutrino telescopes should report
their sensitivity in the future so that telescopes can be meaningfully
compared. Another advantage of following this
proposal is that if a new model arrives long after the collaboration
is still able and willing to run its simulation tools, meaningful
limits can still be extracted via Equation~\ref{nevents}.
(Dramatically different interaction models, for example changing
neutrino cross sections by orders of magnitude, would still
require separate study.)
\begin{itemize}
\item Quote $[A\Delta\Omega]$ and/or $[V^{w.e}\Delta\Omega]$ as a
function of energy  {\it for each neutrino species}, which I
will call here the aperture.   Note that this includes upstream
attenuation of neutrinos by the Earth.
\item Specify which cross sections were used.
\item Specify the livetime $\Delta t$ the observations correspond
to.  Deadtime corrections belong here, not in the quoted aperture.
\item Give the expected background events during 
the livetime  $\Delta t$, if any.
If this depends on the threshold applied, these steps should be repeated
for a few thresholds.
\item Quote the aperture for cross sections 75\% and 125\% the
nominal value used.  If the effect is non-linear, more variations
would be appropriate.
\item Quote the aperture for some variation of the inelasticity,
$y$, as well.
\item Number of events observed, if any.  If significant, this
list should be repeated for a range of detector thresholds and observed
events.
\end{itemize}
With the above information, comparisons could be directly
made without subsequent interpretation.  Also limits
on new models could be set using old data.

\section*{7. Conclusions}

\hs Detection of cosmic neutrinos offers a fertile ground for addressing
interesting questions in astrophysics as well
as elementary particle physics.   The generation of neutrino
telescopes coming online within the next 3--5 years will greatly increase
the existing apertures and probe theoretically interesting regions.
Below $10^{16}$~eV, neutrino astronomy is the domain of the optical
Cherenkov detectors.   Above $10^{17}$~eV, radio-based detectors
offer the largest apertures and are well matched to the 
expected spectrum of GZK neutrinos.    The acoustic technique also shows
promise if ongoing work to demonstrate lower energy thresholds is
successful.   The variety of techniques applied to the challenge
of neutrino astronomy is impressive.   I have presented a proposal
for describing the sensitivity of these widely varying 
techniques in a common framework.

\bigskip\noindent
{\large \bf Acknowledgments}

The author thanks the Nobel Symposium Committee for
its hospitality and organization of an excellent workshop.
This work was partially supported by the U.S. Department
of Energy's Office of Science and the National Aeronautics and Space 
Administration.  I thank the many representatives of the collaborations
mentioned here as well as Amy Connolly,
Bob Cousins, and Jay Hauser for helpful discussions.

\newpage

\begin{table}[!htp]
\begin{center}
\begin{tabular}{|l|l|l|l|}
\hline\hline
term      &   symbol &  with arguments & units for neutrinos\\
\hline\hline
neutrinos &
  $N_{\nu}$    & 
  $N_{\nu}$ &
  (unitless) \\
\hline
neutrino intensity & 
  $I$ &
  $I(\theta,\phi)$  &
  [length]$^{-2}$ [sr]$^{-1}$ [time]$^{-1}$\\
\hline
neutrino brightness, {\it or} & 
  $I_{E}$, &
  $I_{E}(E, \theta, \phi)$ &
  [length]$^{-2}$ [sr]$^{-1}$ [time]$^{-1}$ [energy]$^{-1}$\\
neutrino specific intensity  &
  &
  &
  \\
\hline
net flux & 
  $F_{E}$ &
  $F_{E} (E)$ &
  [length]$^{-2}$ [time]$^{-1}$ [energy]$^{-1}$\\
flux density & & & \\
\hline 
total integrated flux,&
  $F$ &
  $F$ &
  [length]$^{-2}$ [time]$^{-1}$\\
{\it or} surface flux & & & \\
{\it or} flux density & & & \\
\hline\hline
\end{tabular}

\caption{Definitions of relevant quantities used here.  Note
that the term ``flux density'' finds more than one use in
the literature, but the more common is for $F_E$.
\label{defns}}
\end{center}
\end{table}

\begin{table}[!htp]
\begin{center}
\begin{tabular}{|l|l|l|l|l|l|l|}
\hline\hline
\multicolumn{2}{|c|}{~}&\multicolumn{4}{|c|}{$N_{\mbox{events}}$}\\
\hline
\multicolumn{2}{|c|}{~}&
\multicolumn{1}{|c|}{Top. Def.}&
\multicolumn{2}{|c|}{GZK}&
\multicolumn{1}{|c|}{WB}\\
\hline\hline
Telescope & Duration & (PS)
&(min) & (max) &  ~ \\
\hline
Amanda-II & 3 live years &  0.6   & 0.02 & 0.1  & 125 \\
ANITA & 45 live days       & 44 & 4.8 & 18 & 6.5\\
Auger & 3 live years       & 0.7 & 1.0& 3.0 & 1.1\\
EUSO & 2.7 live years    & 18 &  0.9 & 3.6 & 1.9\\
IceCube & 3 live years    & 1.1 & 0.5 & 1.3 & 281\\
RICE & 3 live years      & 3.3 & 0.9 &2.8 & 1.2\\
SALSA-1000 Rx& 3 live years       & 50 & 58 & 194 & 56\\
\hline
\hline
\end{tabular}
\caption{Estimated number of events from various models that
would be observed in some characteristic telescopes which
are being constructed or considered.  
The WB~model
is a $E^{-2}$ spectrum with the Waxman-Bahcall~\cite{wb} coefficient. 
Disclaimer: these numbers include my own estimates of the detector
apertures as described in the text.  A more accurate comparison
could be achieved if more of the experiments follow the proposal
described in the text.  Here the Amanda numbers correspond
to $\nu_\mu$ only; they have also recently reported sensitivity to
other flavors.
\label{predictions}}
\end{center}
\end{table}

\end{document}